# A Deformable Interface for Human Touch Recognition using Stretchable Carbon Nanotube Dielectric Elastomer Sensors and Deep Neural Networks

Chris Larson[1], Josef Spjut[2], Ross Knepper[3], Robert Shepherd[1,4]

**Abstract**
This paper presents a machine learning approach to map outputs from an embedded array of sensors distributed throughout a deformable body to continuous and discrete virtual states, and its application to interpret human touch in soft interfaces. We integrate stretchable capacitors into a rubber membrane, and use a passive addressing scheme to probe sensor arrays in real-time. To process the signals from this array, we feed capacitor measurements into convolutional neural networks that classify and localize touch events on the interface. We implement this concept with a device called OrbTouch. To modularize the system, we use a supervised learning approach wherein a user defines a set of touch inputs and trains the interface by giving it examples; we demonstrate this by using OrbTouch to play the popular game Tetris. Our regression model localizes touches with mean test error of 0.09 mm, while our classifier recognizes gestures with a mean test accuracy of 98.8%. In a separate demonstration, we show that OrbTouch can discriminate between different users with a mean test accuracy of 97.6%. At test time, we feed the outputs of these models into a debouncing algorithm to provide a nearly error-free experience.

## 1. Introduction

Humans and other animals demonstrate a remarkable ability to map sensory information from their skin onto internal notions of hardness, texture, and temperature in order to reason about their physical environment. This capability is enabled by massively parallelized neural computation within the somatosensory cortex, which is fed by a network of nerve cells distributed throughout the epidermis. Recent advances in stretchable electronics, soft robotics, and non-convex optimization methods for deep neural networks now offer us building blocks on which we can start to replicate this tactile perception synthetically. Inspired by biological skins, in this work we have leveraged these advances to develop OrbTouch, a device that interprets tactile inputs using deep neural networks trained on examples provided by a user.

Figure 1 illustrates the OrbTouch concept. We monolithically integrate stretchable carbon nanotube (CNT) capacitors into its rubber membrane to create a soft haptic interface. The sensing apparatus is composed of an overlapping mesh of CNT films, in which orthogonal traces are separated by a thin layer of rubber, forming a parallel plate capacitor at each intersection. Our sensing matrix is designed to enable the independent addressing of $n^2$ sensors using $2n$ electrical connections. To localize interactions on the interface, we feed a single sensor output vector (i.e., from one timestep) into a 2D convolutional neural network (CNN) that regresses the coordinates of touch events. To classify these events, which may vary in abstraction from a simple poke (Fig 1a) to gestures producing complex deformations that evolve over time (Fig 1b,c), we convolve a 3D filter over several timesteps of the incoming data stream to capture the relevant spatiotemporal features. As simple demonstrations of this idea, we use OrbTouch to play the video game Tetris, in real-time at a sampling rate of 10 Hz, and also to identify users.

The remainder of this article is organized as follows: in Section 2 we briefly discuss recent advances in shape changing interfaces, haptics, stretchable sensing, as well as literature from the deep learning and statistical machine learning communities on which our approach is motivated. Section 3 covers the design and fabrication of the OrbTouch device, while Section 4 covers the signal processing architecture, training methods, and training results. Section 5 provides an overview the software implementation and highlights two example applications of OrbTouch. In section 6 we provide a contextual overview of these results and also provide information theoretic analyses of our training data to better understand the information density in our interface, and its potential to be used for more sophisticated functions. Finally, Section 7 concludes the article by briefly discussing future research directions and associated challenges.

---
[1]Department of Mechanical Engineering, Cornell University
[2]NVIDIA Research, NVIDIA Corporation
[3]Department of Computer Science, Cornell University
[4]Department of Materials Science and Engineering, Cornell University

**Corresponding author:**
Robert Shepherd, Cornell University, 553 Upson Hall, Cornell University, Ithaca, NY 14853, USA.
Email: rfs247@cornell.edu

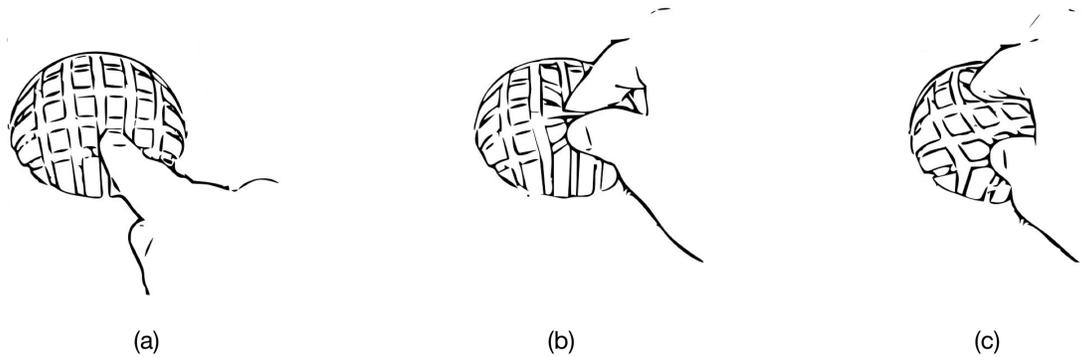

**Figure 1.** Illustration of the OrbTouch concept. A dome shaped balloon is inflated to render a haptic interface, through which a user transmits information by deforming it. Both the syntax and the semantics of the input patterns can be specified by the user. Outputs from an array of capacitors embedded in the membrane are fed through a series of convolutional neural networks trained to localize interactions, such as the finger press shown in (a), as well as recognize abstract events, such as pinching (b) and twisting (c), that evolve over time, yet constitute discrete inputs.

## 2. Related Work

User interfaces provide an interactive window between physical and virtual environments. Traditionally, the tactile interface facilitating this interaction has been capacitive touch screens, keyboard buttons, and the computer mouse. Making physical interaction more rich, both in terms of expanding the type and complexity of inputs that are available to the user, as well as the physical rendering of virtual objects, is of fundamental interest to the fields of human computer interaction (HCI), human robot interaction (HRI), and virtual reality (VR).

Recently, researchers have started to adopt strategies from the field of soft robotics (Rus and Tolley 2015) to augment the touch experience, creating tangible interactions that go beyond tapping, swiping, and clicking. Follmer, Leithinger, Olwal, Cheng and Ishii (2012) used the concept of particle jamming, developed by Brown, Rodenberg, Amend, Mozeika, Steltz, Zakin, Lipson and Jaeger (2010), to create a passive haptic interface that the user can freeform shape and then freeze in place. More recently, Stanley and Okamura (2017) developed an active version of this interface, which dynamically renders 3D topologies using a grid of connected rubber cells controlled by pneumatic inputs, particle jamming, and a spring-mass based kinematic model. Deformable haptic interfaces are a promising area of research with opportunities to leverage microfluidic technologies (Russomanno, OModhrain, Gillespie and Rodger 2015) to enable shape changing interfaces for teleoperations, VR, and braille displays.

In addition to using soft haptic interfaces for physicalization, there are efforts to understand how we can use the passive dynamics of deformable materials (Lee, Kim, Jin, Choi, Kim, Jia, Kim and Lee 2010; Rasmussen, Pedersen, Petersen and Hornbæk 2012), and even the human epidermis (Ogata, Sugiura, Makino, Inami and Imai 2013; Weigel, Mehta and Steimle 2014), as a medium for communication. A significant challenge in this pursuit pertains to sensing finite deformation in the compliant medium, as well as signal processing and software for robust mapping of sensory data to continuous states, for functions such as finger tracking, as well as discrete states to recognize user intent or emotion.

Pai, VanDerLoo, Sadhukhan and Kry (2005) developed a passive haptic ball with embedded accelerometers and an outer enclosure containing flexible capacitors. They used an extended Kalman filter to estimate ball orientation and finger positions using their bi-modal sensor input. Han and Park (2013) created a conceptually similar device and demonstrated the ability to recognize different grips with a classification accuracy of ∼98% using a support vector machine (SVM) classifier. Tang and Tang (2010) developed a dome-shaped foam interface and used Hall-effect sensors positioned around the base of the interface to capture a set of predefined interactions. In perhaps the most simple approach, Nakajima, Itoh, Hayashi, Ikeda, Fujita and Onoye (2013) placed a microphone and a barometer inside of a balloon and were able to discriminate grasps, hugs, punches, presses, rubs, and slaps with a mean classification accuracy of 81.4% using an SVM classifier. Vision based sensing has also been explored. Harrison and Hudson (2009) used an infrared camera, placed behind the interface to capture a bottom-up view of the deforming membrane, in conjunction with blob detection algorithms to localize touch interactions. Other researchers have used vision with different interface designs (Steimle, Jordt and Maes 2013). Although vision based sensing is inherently high dimensional and sensitive to deformation, focal length and camera placement impose two very significant constraints on the system design.

Both the human somatosensory system and capacitive touch displays alike benefit from high dimensional tactile sensory input. It is our view that, by embedding sensors directly into the touch surface, we will similarly enable the widest range of functional soft interface designs. To accomplish this, we can leverage stretchable electronics (Rogers, Someya and Huang 2010), which has enabled new capabilities across many applications such as *in vivo* biosensing (Viventi, Kim, Vigeland, Frechette, Blanco, Kim, Avrin, Tiruvadi, Hwang and Vanleer 2011), robotics (Kim, Lee, Shim, Ghaffari, Cho, Son and Jung 2014), and soft robotics (Larson, Peele, Li, Robinson, Totaro, Beccai, Mazzolai and Shepherd 2016). Charge conduction in stretchable media can be achieved using many different strategies, such as back filling channels embedded in elastomers with low $T_m$ liquid eutectic alloys (Park, Majidi, Kramer, Bérard and Wood 2010) or ionically conducting hydrogel polymers (Keplinger, Sun, Foo, Rothemund, Whitesides and Suo 2013), depositing silicon thin films with serpentine patterns to enable them to stretch by uncoiling (Khang, Jiang, Huang and Rogers 2006), and using carbon nanotubes (CNTs) (Yamada, Hayamizu, Yamamoto, Yomogida, Izadi-Najafabadi, Futaba and Hata 2011). Lipomi, Vosgueritchian, Tee, Hellstrom, Lee, Fox and Bao (2011) recently made transparent electrode films that remain conductive to within one order of magnitude by aerosol spraying a dilute suspension of CNTs in N-methylpyrrolidone onto a PDMS substrate. This combination of high conductivity at high strains, coupled with ease of fabrication makes CNTs an excellent choice for shape changing user interfaces.

In additional to improved sensing methods, there is a simultaneous need for robust signal processing architectures that are suited for stretchable electronics. As evidenced by recent trends in computer vision and deep learning (LeCun, Bengio and Hinton 2015), enabling tactile sensing machinery to reason about the physical world in a meaningful way will likely require high-capacity models that learn from data efficiently. This is important for emerging touch sensing methods in VR (Shepherd, Peele, Murray, Barreiros, Shapira, Spjut and Luebke 2017), wearable sensing (Stoppa and Chiolerio 2014), HRI (Hughes, Lammie and Correll 2018), and HCI (Roh, Hwang, Kim, Kim and Lee 2015) that are being used for increasingly complex recognition tasks. Systems based on deep neural networks have surpassed, or are approaching, human capabilities in a number of areas including the classification and segmentation of both natural and medical images (He, Zhang, Ren and Sun 2016), playing Atari games (Mnih, Kavukcuoglu, Silver, Graves, Antonoglou, Wierstra and Riedmiller 2013), playing high complexity board games *silver2016mastering, interpreting natural language (Mikolov, Chen, Corrado and Dean 2013), and sequence recognition (Bengio, Vinyals, Jaitly and Shazeer 2015). Artificial neural networks are known for their representational power, and convolutional filtering is particularly suited for inputs that are spatially or temporally correlated.

For embedded sensing in shape changing interfaces, we note that the laws of material continuity and elasticity dictate that resistive and capacitive elements in a simply connected body must converge in value as the distance between them approaches zero. Sensors distributed throughout a simply connected body, therefore, are inherently spatially correlated. This observation motivates our modeling approach.

## 3. OrbTouch

Our shape changing interface, OrbTouch (Fig. 2a), consists of a pressurized silicone orb with an embedded array of stretchable CNT capacitors. Each CNT electrode is bonded to an external Cu lead that is routed through an analog-digital converter (ADC) to the GPIO interface on a Raspberry Pi 3 (RBPI3; Fig. 2b). To train the device, there is a push-button adjacent to the interface that the user presses during training to supplement the logged data with ground-truth labels. Models are trained offline and then uploaded onto the RBPI3, which computes them directly in the sensor measurement loop in real time. In addition to computing neural networks, we use the RBPI3 to control the sensing peripherals as well as host communication via bluetooth.

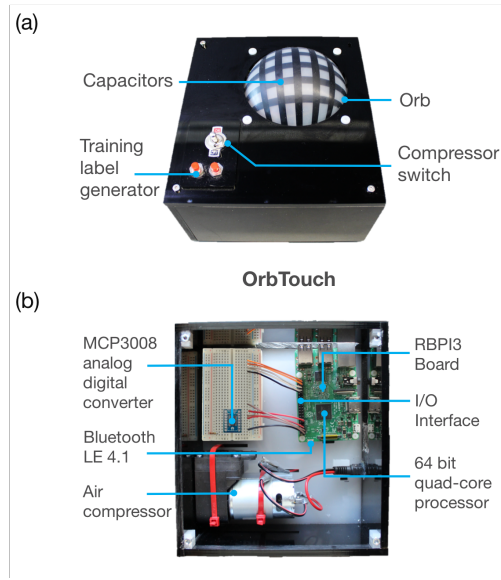

**Figure 2.** Photographs of the OrbTouch device. (a) Its embedded capacitors capture shape changes caused by human touch. (b) The internal components of OrbTouch consist of an embedded RBPI3 computer, ADC, and air compressor used to control pressure in the orb.

### 3.1 Sensor Fabrication

Figure 3 shows the internal construction and configuration of the CNT-DESs and OrbTouch membrane. Each sensor consists of a parallel plate capacitor with two blended MWCNT-SWCNT thin film electrodes separated by a PDMS dielectric layer. The electrodes are patterned by aerosol spraying a dispersion of the CNTs in a solution of 2-propanol and toluene (adapted from Lipomi, Vosgueritchian, Tee, Hellstrom, Lee, Fox and Bao 2011) through a stencil on the base PDMS substrate.

Our process is performed in several steps: (i) in a beaker, a blended mix of MWCNT (P/N 724769; Sigma Aldrich Corp.) and SWCNT (P/N P3-SWNT; Carbon Solutions Inc.) are dispersed in a 90/10 (vol. %) solution of 2-propanol (P/N 278475, Sigma Aldrich Corp.) and toluene (P/N 244511; Sigma Aldrich Corp.) at a concentration of 0.05 wt.% using a centrifugal mixer (SR500, Thinky U.S.A. Inc.) in combination with ultrasonic agitation. (ii) A ∼0.5 mm layer of silicone rubber (Ecoflex-0030, Smooth-on Corp.) is cast onto an acrylic sheet and cured. (iii) A layer of tape is overlaid onto the substrate and a laser cutter (Zing 24, Epilog Laser Corp.) is used to selectively remove portions of it to form the bottom electrode pattern. (iv) The CNT dispersion is sprayed through the mask with an airbrush (eco-17 Airbrush Master, Master Inc.) to form the bottom electrode. Several coats are applied until each trace reaches an end-to-end resistance of ∼1 kohm. (v) The mask is then removed and a thin (∼0.5 mm) dielectric layer (Ecoflex-0030) is cast over the entire substrate and cured. (vi) Steps iii-v are repeated (in reverse order) to form the top half of the membrane (overall thickness ∼2 mm). (vii) External Cu leads are attached to each of the 10 CNT electrodes and connected to the ADC and RBPI3.

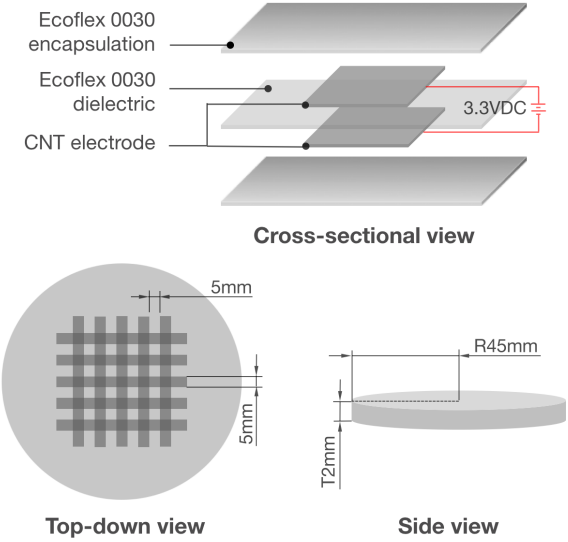

**Figure 3.** Membrane and sensor architecture. The interface is composed of upper and lower PDMS encapsulation layers, upper and lower CNT film electrodes, and a 0.5 mm PDMS dielectric layer, yielding a total thickness of ∼2mm. The sensors are configured into a passive matrix, where each electrical lead in the grid measures 5×55 mm, yielding an overall density of 1 sensor cm$^{-2}$.

### 3.2 Sensing Method

The sensing grid is designed as a passive matrix that enables us to position 25 sensors over the surface using only 10 electrical connections. To measure capacitance we use the digital I/O pins on the RBPI3 and an ADC. To isolate the $i, j^{th}$ sensor, where $i, j \in \{0, 1, 2, 3, 4\}$, we set the $i^{th}$ electrode to +3.3VDC (vertical orientation, Fig. 4a), and monitor the corresponding voltage change on the $j^{th}$ electrode (horizontal orientation, Fig. 4a), with the remaining electrodes connected to ground on the RBPI3 chassis to reduce cross talk and interference. Figure 4b shows the equivalent circuit of the measurement. The capacitance in our sensor grid is 41.2 pF (SD = 2.9 pF). We use a 50 Mohm resistor to achieve a nominal RC time constant of $\tau_0$ ∼2 ms. When the $i, j^{th}$ sensor is being measured, the $i^{th}$ column electrode is set to +3.3 VDC while the $j^{th}$ row electrode, which is routed through the ADC, is disconnected from ground. A second capacitor (1 pF) is placed in series with the $j^{th}$ row electrode and the ADC to shift the polarity of $V_m$ into the 0-3.3V range for the RBPI3.

### 3.3 Deformation-Capacitance Model

The sensors in OrbTouch behave according to the parallel plate capacitance formula, $C \propto A/d_t$, where $C$ is the capacitance of the sensor, $A$ is the surface area of the sensor, and $d_t$ is the dielectric thickness. To validate this experimentally, we develop a simple model of capacitance for incompressible, inflating shells, and compare its predictions to measured values that we obtain by inflating the interface.

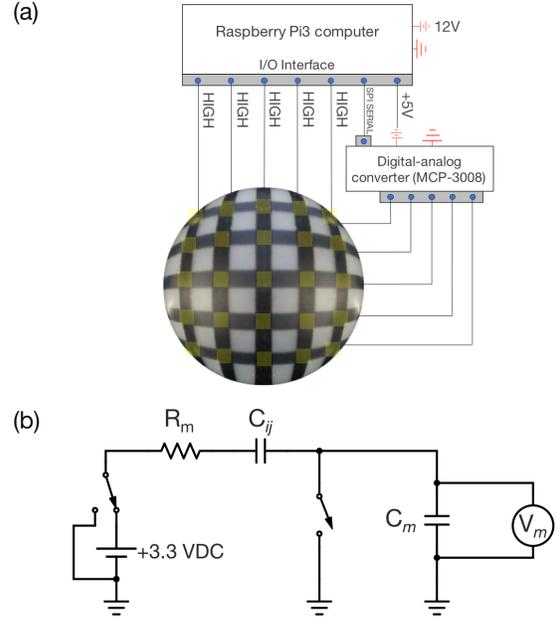

**Figure 4.** Capacitance measurement method. (a) To measure capacitance, we set one vertical electrode HIGH (+3.3 VDC) and monitor the induced voltages on the orthogonal electrodes using an ADC, which relays the signals to the RBPI3 over SPI serial. During each measurement, there is one pin set HIGH, and one pin that is read; the remaining eight electrodes are connected to ground to minimize cross talk between neighboring electrodes and electromagnetic interference. (b) Equivalent measurement circuit. The $i, j^{th}$ capacitor is represented by $C_{i,j}$. The nominal capacitance of our sensors is 41.2 pF (SD = 2.9 pF). We use a $R_m$ = 50 Mohm inline resister to yield an RC time constant of $\tau$ ∼2 ms. We use a second capacitor, $C_m$ = 1 pF, to flip the polarity of the measured ($V_m$) voltages.

We first define three principle stretches, $\lambda_1, \lambda_2, \lambda_3$, using a Cartesian basis as shown in Figure 5a. In an incompressible (i.e., $\lambda_1\lambda_2\lambda_3 = 1$) rubber dielectric under equibiaxial tension (i.e., $\lambda = \lambda_1 = \lambda_2$), the fractional change in capacitance is a function of only its radial stretch,

$$\begin{aligned}\frac{C}{C_0} &= \frac{\lambda_1\lambda_2}{\lambda_3} \\ &= \lambda_1^2\lambda_2^2 \\ &= \lambda^4\end{aligned} \quad (1)$$

Because it is difficult to measure $\lambda$ experimentally, we derive an alternative to (1) that depends on the membrane deflection, $d_{def}$ (Fig. 5a), which we can measure, using the well known approximation,

$$A_{orb} \approx \left[\frac{r^{16/5} + 2\left(rd_{def}\right)^{8/5}}{3}\right]^{5/8}, \quad (2)$$

that expresses the surface area of the hemispheroidal orb, $A_{orb}$, in terms of its radius, $r$, and $d_{def}$. If we assume that the deformation is homogeneous over the entire membrane as it inflates, we can alternatively express the quartic stretch term as $\lambda^4 = \left(A_{orb}/A_{orb,0}\right)^2$, where the nominal surface area is simply given by $A_{orb,0} = \pi r^2$. Combining these expressions with (2) yields the desired relationship between fractional change in capacitance and $d_{def}$.

$$\frac{C}{C_0} \approx 4\left(\frac{1}{3} + \frac{2}{3}\left(\frac{d_{def}}{r}\right)^{8/5}\right)^{5/4}. \tag{3}$$

Figure 5(b) plots the mean capacitance of our 5×5 capacitor grid versus our parameterized function, $\lambda^4(d_{def}, r)$, under controlled inflation. The observed behavior undershoots our prediction; this has been observed previously (*cf.* Keplinger, Sun, Foo, Rothemund, Whitesides and Suo 2013), and is commonly attributed to a decrease in dielectric permittivity that occurs in elastomers as they are stretched. We also note two other potential sources of error, the first being our approximation of the orb as a hemispheroid (*cf.* Adkins and Rivlin, 1955). Secondly, we assume that the deformation in the orb is homogeneous, however, sensors near the perimeter of the membrane are closer to the clamped boundary and therefore deform differently than sensors near the center. Although we use a simplified model, the general relationship between capacitance and quartic stretch is quasi-linear, as predicted. We also note that each sensor in the grid is well defined, varying monotonically with the quartic radial stretch. This behavior suffices for our application, as we use these sensors to learn latent representations of deformation with neural networks, not for explicit shape estimation.

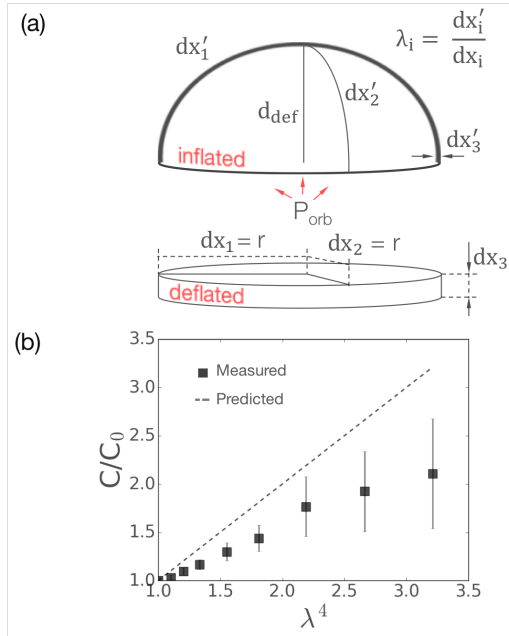

**Figure 5.** Relationship between deformation and capacitance in the orb. (a) Free body diagram of the touch membrane in the undeformed (deflated) and deformed (inflated) states. Under inflation we assume equibiaxial tension, and thus, because the membrane is incompressible, its stretch state is fully described by the radial stretch. (b) Plot of $C/C_0$ versus $\lambda^4$ ($n$ = 25).

## 4. Computational Approach

### 4.1 Model Architecture

Our signal processing architecture is designed for modular touch interaction, enabling one to fully define both the syntax and semantics of a set of inputs for a given application. We build this capability on top of two core functions: gesture recognition and touch localization, both of which are implemented using light-weight convolutional neural networks. As inputs to our models, we use sensor images that are computed as follows: $z := C/C_0$ ($z \in \mathbb{R}^{5 \times 5}$), where $C_0$ is the mean baseline capacitance taken over a 10 second interval at the beginning of each session. For gesture recognition, we use an inference model based on a 3D-CNN (*CNN-3D*), to map a queue of $m$ sensor images, $z_0 : z_9$, to a categorical probability distribution, $p_c$, over $n_c$ gesture classes ($\mathbb{R}^{5 \times 5 \times 10} \to \mathbb{R}^{n_c}$). We use CNN-3D to identify gestures, and also to discriminate between users performing the same gesture. For touch localization we use a regression model (*CNN-2D*) that uses 2D convolutions, which map sensor readings, $z$, from one timestep to a continuous $d$-dimensional space ($\mathbb{R}^{5 \times 5} \to \mathbb{R}^d$). We use CNN-2D to estimate touch location on the curvlinear surface (i.e., $d = 2$), however, it could also be used to estimate membrane deflection, touch pressure, or other continuous quantities.

Figure 6 shows the architectural features of the CNN-3D and CNN-2D models. CNN-2D convolves its kernels over the spatial dimensions of the input, whereas CNN-3D convolves 3D kernels over the spatial and temporal dimensions in order to capture the dynamics of the touch gesture. Expression (4) provides an algebraic representation of the convolutions in these networks,

$$\begin{aligned} a^l_{mijk} &= \sigma\left(w^l_m * \mathbf{a}^{'l-1} + \mathbf{b}^l_m\right)_{ijk} \\ &\leftarrow \sigma\left(b^l_m + \sum_n \sum_q \sum_r a^{'l-1}_{i-n, j-q, k-r} w^l_{mnqr}\right), \end{aligned} \tag{4}$$

where $a^l_{mijk}$ refers to the $(i, j)^{th}$ node in the $m^{th}$ feature map in layer $l$, $w^l_m$ and $b^l_m$ are the convolutional kernel and bias terms corresponding to the $m^{th}$ feature map in layer $l$, respectively, the operator $*$ denotes the convolution between the kernel and its input, and $a^{',l-1}$ represents the zero-padded input to layer $l$ (we employ *same* padding). Expression (4) is valid for both the 2D and 3D convolutions (the time dimension, indexed by $k$, in the CNN-2D is singleton). The dense layers following the convolutional layers are mapped using the inner product of the weight matrices with the nodes from the preceding layer. CNN-3D uses a softmax activation in its output to produce a probability over gesture classes, whereas CNN-2D uses a *tanh* activation to regress continuous valued coordinates of touch. Both networks use interior rectified linear unit (*relu*) activations.

To run these models on the RBPI3 in real-time, we had to consider trade-offs between model depth, number of timesteps in the input, $t$, and sampling rate, $\omega$. Ideally we would use deep models in combination with a high-bandwidth input, however, we cannot simultaneously maximize model depth, $t$, and $\omega$ in our compute- and time-constrained system. Through observing different users, we noticed that touch gestures are typically ∼1 second in duration. Using $t\omega^{-1} = 1s$ as a constraint, we found that a window of $t = 10$ and a sampling rate of $\omega = 10$ s$^{-1}$ allows us to capture the relevant features from gestures. To enable the system to run safely at a latency of < 100 ms, we use relatively shallow neural networks each with two convolutional layers and two fully connected layers.

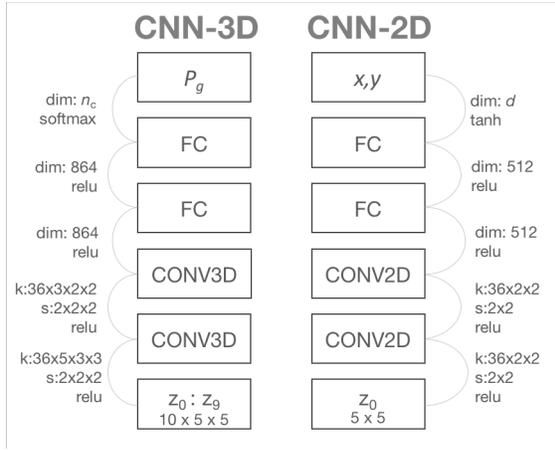

**Figure 6.** Computational graph of the inference (CNN-3D) and regression (CNN-2D) models. Both networks have two hidden convolutional layers and two hidden fully connected layer. The kernel size, $k$, and stride, $s$, of each convolutional operation are provided. Network CNN-3D accepts as input a sliding window of 10 discrete sensor readings (10×5×10) and outputs a probability distribution over $n_c$ classes using a softmax activation on the output. Because the information in a gesture is inherently spatiotemporal, we convolve a 3D kernel over both the spatial and temporal dimensions of the input to capture relevant features. Network CNN-2D accepts a 5×5 sensor matrix and outputs a continuous valued vector using a tanh activation on the output layer.

### 4.2 Optimization Methods and Training Results

We teach OrbTouch new inputs by pressing the label button, located adjacent to the orb (Fig. 2a), in unison with the imparted gesture. The label button is connected to the I/O interface on the RBPI3 computer, and its state is logged at every time step. We optimize models CNN-3D and CNN-2D stochastically on the logged data using an external computer, and then upload the trained parameters back onto the RBPI3 to use the device as a touch controller. To demonstrate this process, we define a set of five simple inputs: a finger press, a clockwise twisting motion, a counterclockwise twisting motion, a pinching motion, and a null input. We collected ∼5 min of labelled training data for each of the above input classes, yielding $n = 1.75 \times 10^4$ total examples. The parameters in CNN-3D are optimized using the categorical cross-entropy loss, $\ell_{CE}$ (5), with 2-norm regularization applied to its weights, where $l$ indexes the layers in the network, and $m$ indexes the feature maps in layer $l$. We used mini-batches of $n = 150$, and regularization constants $\lambda_{CE1} = 5 \times 10^{-4}$, $\lambda_{CE2} = 1 \times 10^{-5}$. Optimization was implemented using the adaptive momentum estimation algorithm (ADAM) from Kingma and Ba (2014).

$$\ell_{CE}(h(z)) = -\frac{1}{n}\sum_{i=0}^{n}(y\ln(h(z)) + (1-y)\ln(1-h(z)))$$
$$+ \lambda_{CE1} \sum_{l=1}^{2} \sum_{m=0}^{M} ||w_m^l||_2^2$$
$$+ \lambda_{CE2} \sum_{l=3}^{5} ||w^l||_2^2, \quad (5)$$

We performed all training offline on a single GPU (GeForce GTX 1080 Ti, NVIDIA Corp.) using the Tensorflow framework (Abadi, Agarwal, Barham, Brevdo, Chen, Citro, Corrado, Davis, Dean and Devin 2016). Figure 7a plots the training and validation accuracy of CNN-3D versus training epoch. CNN-3D reaches a a test accuracy of ∼98.8% following ∼500 epochs. Figure 7b plots the learning curve between this model and dataset, indicating that the model achieves >95% classification accuracy using $5 \times 10^3$ examples, which is the equivalent of ∼ 10 minutes of training.

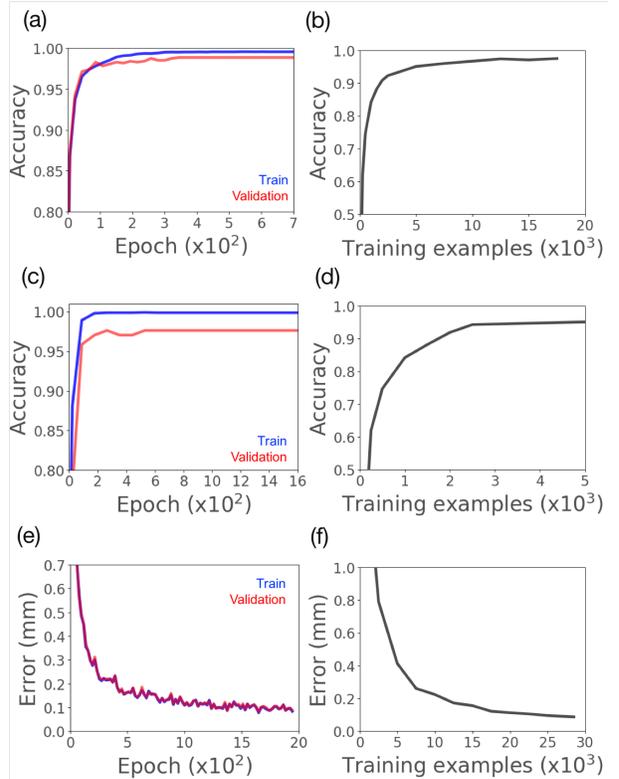

**Figure 7.** CNN training results. (a) Plot of binary classification accuracy versus training epoch for CNN-3D on the gesture recognition dataset. We measure a test accuracy of 98.8% after $5 \times 10^2$ epochs ($n = 1.75 \times 10^4$). (b) Learning curve of CNN-3D on the gesture recognition dataset. (c) Plot of binary classification accuracy versus training epoch for CNN-3D on the user identification dataset. We measure a test accuracy of 97.6% after $6 \times 10^2$ epochs ($n = 5 \times 10^3$). (d) Learning curve of CNN-3D on the user identification dataset. (e) Plot of the mean absolute error of CNN-2D on the touch location dataset, measured in mm, over $2 \times 10^3$ epochs ($n = 2.85 \times 10^4$). (f) Learning curve of CNN-2D on the touch location dataset.

In addition to gesture recognition, we also trained CNN-3D to identify, from a set of $n_c = 10$ users, the person interacting with the device. In this experiment, each participant performed the clockwise twisting motion, as defined previously, for ∼5 min. We then trained CNN-3D using similar hyperparameters to those used for the gesture recognition data, achieving a test accuracy of 97.6% (Fig. 7c). Figure 7d plots the learning curve for this dataset. We observe only a marginal decrease in test accuracy on the user recognition data set despite its larger number of output classes ($n_{c,user} = 10$ vs $n_{c,gesture} = 5$) and much more nuanced differences between the $n_{c,user}$ classes. In both cases, we believe our model capacity is limited primarily by our manual labeling method, which introduces noise into our response variable due to non-uniform shifts between ground truth labels and the imparted gestures.

To train the CNN-2D model, we had a user visually locate the sensors on the membrane and press them (on, off) for a total of ∼30 min ($n = 1\times10^4$). We use ridge regression (6) to optimize the parameters in CNN-2D using the Nesterov accelerated gradient (NAG) algorithm from Nesterov (1983). Figure 7e plots mean absolute error (MAE) versus training epoch; we achieve a test error of MAE = 0.09 mm, while Figure 7f plots the learning curve for this dataset. Our best convergence and training performance was achieved using mini-batches of $n = 128$, gradient clipping ($||\nabla_{global}||_2 \leq 10.0$), regularization constants $\lambda_{MSE1} = 1\times10^{-5}$, $\lambda_{MSE2} = 5\times10^{-6}$, and by adding zero-mean Gaussian noise (SD = 0.05 mm) to each ground truth label. For simplicity, we report distances with respect to the undeformed membrane that lies in two-dimensions (i.e., its circular state), where the touch surface spans the x-y interval [(0, 0), (4,4)] mm. Thus, for a membrane deflection of $d_{def} = r$, a multiplicative factor of $\pi/2$ provides an approximation of the true error along the curvilinear surface of the orb.

$$\ell_{MSE}(h(z)) = -\frac{1}{n}\sum_{i=0}^{n}(y - h(z))^2$$
$$+ \lambda_{MSE1}\sum_{l=1}^{2}\sum_{m=0}^{M}||w_m^l||_2^2 \qquad (6)$$
$$+ \lambda_{MSE2}\sum_{l=3}^{5}||w^l||_2^2,$$

## 5. Tetris

To demonstrate how these models can be integrated into software applications, we use OrbTouch to play the popular video game Tetris (Fig. 9a-e). The objective of Tetris is to place a random cascade of falling pieces, or *Tetrominos*, into a bounding rectangle without filling it up; filling a row causes the Tetrominos in that row to disappear, allowing the pieces above it to drop and thus preventing the game board from filling. During gameplay, we use OrbTouch to translate (Fig. 9b,e) and rotate (Fig. 4c,d) the Tetrominos as they fall using the gestures that we defined in Section 4. We implement this with a C++ program running on the RBPI3, which executes sensor measurements, neural network computation, and Bluetooth communication with the host (Fig. 9f). We enqueue sensor measurements into a 1 second memory buffer, which gets passed to CNN-3D and CNN-2D at each timestep. The user's gestures are recognized by computing $argmax(p_g)$. When a finger-press is predicted, CNN-2D is used to estimate the location of touch, from which an appropriate translation is generated. Because the output from CNN-3D is noisy (error rate ∼1%), during game play we pass it through a secondary debouncing filter, which in turn relays commands asynchronously to the host.

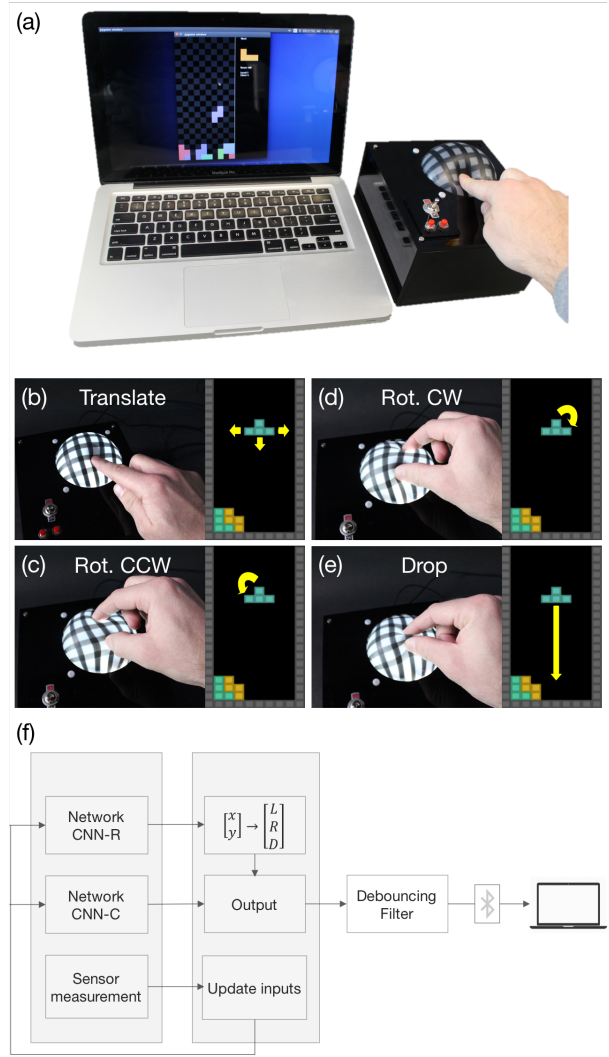

**Figure 8.** Application of OrbTouch to the popular game Tetris. (a) Photograph of OrbTouch being used to control an adaptation of the game Tetris. (b) Finger pressing, or poking are used to translate the Tetromino left, right, and down (L,R,D). (c) Pinching is used to drop the Tetromino directly to the bottom of the grid. (d) Clockwise rotation, or twisting, is used to rotate the Tetromino 90 deg in the clockwise direction. (e) Counterclockwise rotation is used to rotate the Tetromino 90 deg in the counterclockwise direction. (f) OrbTouch software diagram. The first processing step executes capacitance measurements, CNN-2D, and CNN-3D, while the second step generates a command and updates the model inputs for the next timestep. Each of these steps is multithreaded. We use debouncing filter prior to sending commands to the host (via Bluetooth). Each cycle of compute takes ∼86 ms, which fits within our 100 ms target.

Movie 1[*] shows a person performing a random sequence of the Tetris gestures, along with the real-time output of CNN-3D (trained on the gesture recognition dataset). We achieve nearly error-free gesture recognition with OrbTouch using CNN-3D in combination with the debouncing filter. This system runs at a controlled latency of 100 ms, which could be decreased significantly through the use of a GPU.

Movie 2[†] shows a recording of a Tetris game, in which both CNN-3D and CNN-2D are used to generate game commands. The game is controlled using finger-presses (Fig. 8b) to translate the Tetromino (left, down, right), pinching (Fig. 8e) to drop the Tetromino directly to the bottom of the board, clockwise twisting (Fig. 8d) to rotate the Tetromino 90 deg in the clockwise direction, and counterclockwise twisting (Fig. 8c) to rotate the Tetromino 90 deg in the counterclockwise direction. The OrbTouch controller runs as a standalone device, and wirelessly communicates with our Tetris application (written in Python) which runs externally on a laptop computer.

## 6. Information Theoretic Analysis of Sensor Signals

Tetris commands only require $\log_2(7)$ = 2.81 bits of information to encode, which raises the question of whether OrbTouch is capable of encoding more interesting vocabularies of higher perplexity. The performance of CNN-3D on the user identification dataset ostensibly indicates a lower bound of $\log_2(10)$ = 3.32 bits of information in our multivariate sensor signal, however, to gain a more complete understanding of its theoretical limits we consider the complexity of the sensor signals. We evaluate the information content by computing the Shannon entropy, $H(z)$,

$$H(z) = \sum_{i=1}^{n} p(z_i) \log_2 \left( p(z_i) \right) \quad (7)$$

and mutual information, $I(z, y)$,

$$I(z, y) = \sum_{i=1}^{n} \sum_{j=1}^{n} p(z_i, y_j) \log_2 \left( \frac{p(z_i, y_j)}{p(z_i)p(y_j)} \right), \quad (8)$$

of the capacitance data, $z$, and labels, $y$, in the gesture recognition dataset ($n$ = 34,795), where $p(z)$ and $p(z, y)$ represent the marginal and joint probability masses, respectively. To compute $p(z)$ and $p(z, y)$, we first project the data and labels onto the interval [0,1] using min-max normalization, $z \leftarrow (z - z_{min})/(z_{max} - z_{min})$, for each sensor-gesture combination in the dataset, and then concatenate the data for each sensor into a vector of length 34,795. The data and labels are then quantized into 25-bin histograms.

Figure 9 shows a bar chart of the $H(y)$, $H(z)$, and $I(z, y)$ statistics. The complexity of our response variable can be interpreted as follows. Relative to the maximum entropy case in which all five of our gesture classes occur in equal proportion, i.e., $H(y \sim Unif.) = \log_2(5) = 2.32$ bits, the complexity of our response variable is significantly lower, $H(y)$ = 1.28 bits. We expect this given the disproportionate number of static labels in the gesture identification dataset ($p_{g,static}$ = 0.57). On the other hand, we compute a mean signal entropy of $H(z)$ = 2.71 bits, averaged over the 25 sensor channels, indicating that each sensor in OrbTouch contains a surplus of information relative to $y$. Thus, given near optimal encoding of our signal, we theoretically could play Tetris using only one of our sensors. We also use this data to compute the relative entropy between the response variables and covariates, which is a measure of the decrease in uncertainty (in bits) of our response when it is conditioned on the input $z$. We observe a relatively low mutual information, $I(z, y)$ = 0.13 bits, which tells us that although our per-sensor signal entropy is high relative to our response, not all of that information is predictive of the response.

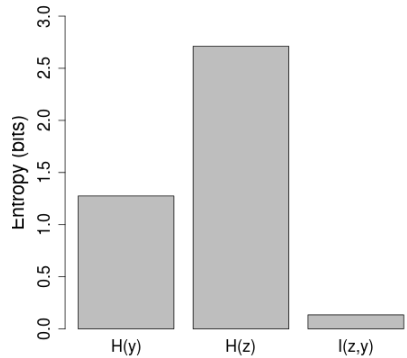

**Figure 9.** Bar chart containing information entropy statistics of the gesture recognition dataset. This dataset consists of 34,795 examples with five categorical labels. The Shannon entropy of a uniformly distributed response variable is $H(y \sim Unif.)$ = 2.32 bits. Here we measure $H(y)$ = 1.28 bits, which is due to the disproportionate number of static labels in the data ($p_{g,static} \sim 0.57$). We measure $H(z) \sim 2.71$ bits averaged over the 25 sensors; significantly higher than the encoding length required for our Tetris game. The low mutual information between our sensors and labels, $I(z, y)$ = 0.13 bits, is indicative of the surplus of information in our signal relative to our response variable. These statistics were computed in R using the *Entropy* package.

While these statistics are computed on time series from individual sensors, the multivariate entropy and mutual information, taken over the 250 dimensional input of CNN-3D, would provide a better estimate of the information that is available to our classifier. Due to the curse of dimensionality, however, estimating the multivariate probability masses is computationally intractable using our quantization method. The effects of spatial and temporal correlation in this data also make it difficult to estimate the true information content in the multivariate signal using these univariate and bivariate statistical measures. In future work, we intend to explore more advanced estimation methods, such as MCMC sampling, to better understand the information in our system, and also to inform better sensor and signal processing design. The high per-sensor entropy in our gesture recognition data (2.71 bits), though, is a promising step towards being able to encode large, interesting vocabularies using deformable interfaces with high-density sensor arrays.

---

[*]https://youtu.be/lStMXNRmRqU
[†]https://youtu.be/82p35vj6M2A

## 7. Discussion

This article explores the use of deformation in a compliant touch surface as a medium for communication. To demonstrate this concept we present OrbTouch, a device that can learn multitouch inputs and localize finger presses, akin to a capacitive touch screen, but one that interprets shape change rather than finger movements. This is enabled by stretchable CNT-based capacitors that we embed inside of the touch surface to provide real-time shape feedback. Rather than use physical models to map sensor data to explicit representations of shape, we leverage deep neural networks, which learn latent representation of deformation, to directly map sensor signals to virtual states that a user can define for their application.

The core of our approach lies in our use of 3D convolutions to capture spatiotemporal features in the gestural inputs. We initially considered other approaches to capture temporal information, including long-short term memory (LSTM) units, as well as CNN feature extractors in combination with LSTMs (Ordóñez and Roggen 2016), however we found that gestures, and even short sequences of gestures, occur over relatively short time horizons. Our approach therefore is to expand the dimension of the input to encompass the relevant time horizon while retaining its spatial and temporal structure, and to use finite impulse response filters to capture the relevant spatial and temporal features. In the future, though, we are interested in expanding the gestural vocabulary to include longer sequences of inputs, which will require the use of recurrent models to capture contextual information.

OrbTouch highlights the utility of statistical approaches and learning algorithms in the rapidly expanding fields of stretchable electronics and soft robotics, and how they can be applied to human computer interaction. Previous research in shape changing interfaces, as well as stretchable electronics, has explored the use of machine learning for sensory mapping. To our knowledge, however, we have demonstrated for the first time the use of stretchable sensors to control a software application in real time. We emphasize the distinction between achieving high performance metrics on in-sample data, for which it is very easy to overfit, and demonstrating that the model generalizes to a real-time data feed such that it can be used to accomplish tasks. This is immensely important in this areas of research, because many of the commonly used stretchable sensors exhibit hysteresis, non-stationarity, and high failure rates.

Although we focus on touch control for human computer interfaces, we believe this approach can also be applied more generally in robotics. OrbTouch's skin could, for example, be overlaid onto a robot and integrated into its perception system; a step towards the level of sensor fusion that we observe in biological systems. A nearer term ambition would be incorporating the skin into robotic end effectors, such as a jamming gripper (Amend, Brown, Rodenberg, Jaeger and Lipson 2012), for robust identification and characterization of grasped objects. Further, in robotics it is generally desirable to have higher dimensional sensing. We designed OrbTouch with 25 sensors, at a density of 1 cm$^{-2}$; however, this choice was motivated by our application and fabrication method. Decreasing the CNT electrode width to 500 $\mu$m using commercially available inkjet printers (Kordás, Mustonen, Tóth, Jantunen, Lajunen, Soldano, Talapatra, Kar, Vajtai and Ajayan 2006), for example, would yield 100 sensors cm$^{-2}$. With a mean per-sensor entropy of 2.71 bits, skins that can sense at this resolution will be an important step towards improving physical perception in robots that use compliant materials.

The code and model parameters used in OrbTouch are available at Larson (2018).

## Acknowledgements

We thank K. O'Brien, B. Peele, K. Petersen, and C.W. Larson for their comments, discussions, and insight. This work was supported by the Army Research Office (grant no. W911NF-15-1-0464) and the Air Force Ofce of Scientic Research (award no. FA9550-15-1-0160).

## References

Abadi, M., Agarwal, A., Barham, P., Brevdo, E., Chen, Z., Citro, C., Corrado, G. S., Davis, A., Dean, J. and Devin, M. (2016), 'Tensorflow: Large-scale machine learning on heterogeneous distributed systems', *arXiv preprint arXiv:1603.04467* .

Amend, J. R., Brown, E., Rodenberg, N., Jaeger, H. M. and Lipson, H. (2012), 'A positive pressure universal gripper based on the jamming of granular material', *IEEE Transactions on Robotics* **28**(2), 341–350.

Bengio, S., Vinyals, O., Jaitly, N. and Shazeer, N. (2015), Scheduled sampling for sequence prediction with recurrent neural networks, *in* 'Advances in Neural Information Processing Systems', pp. 1171–1179.

Brown, E., Rodenberg, N., Amend, J., Mozeika, A., Steltz, E., Zakin, M. R., Lipson, H. and Jaeger, H. M. (2010), 'Universal robotic gripper based on the jamming of granular material', *Proceedings of the National Academy of Sciences* **107**(44), 18809–18814.

Follmer, S., Leithinger, D., Olwal, A., Cheng, N. and Ishii, H. (2012), Jamming user interfaces: programmable particle stiffness and sensing for malleable and shape-changing devices, *in* 'Proceedings of the 25th annual ACM symposium on User interface software and technology', ACM, pp. 519–528.

Han, S. and Park, J. (2013), Grip-ball: A spherical multi-touch interface for interacting with virtual worlds, *in* 'Consumer Electronics (ICCE), 2013 IEEE International Conference on', IEEE, pp. 600–601.

Harrison, C. and Hudson, S. E. (2009), Providing dynamically changeable physical buttons on a visual display, *in* 'Proceedings of the SIGCHI Conference on Human Factors in Computing Systems', ACM, pp. 299–308.

He, K., Zhang, X., Ren, S. and Sun, J. (2016), Deep residual learning for image recognition, *in* 'Proceedings of the IEEE conference on computer vision and pattern recognition', pp. 770–778.

Hughes, D., Lammie, J. and Correll, N. (2018), 'A robotic skin for collision avoidance and affective touch recognition', *IEEE Robotics and Automation Letters* .

Keplinger, C., Sun, J.-Y., Foo, C. C., Rothemund, P., Whitesides, G. M. and Suo, Z. (2013), 'Stretchable, transparent, ionic conductors', *Science* **341**(6149), 984–987.

Khang, D.-Y., Jiang, H., Huang, Y. and Rogers, J. A. (2006), 'A stretchable form of single-crystal silicon for high-performance electronics on rubber substrates', *Science* **311**(5758), 208–212.

Kim, J., Lee, M., Shim, H. J., Ghaffari, R., Cho, H. R., Son, D. and Jung, Y. H. (2014), 'Stretchable silicon nanoribbon electronics for skin prosthesis', *Nature communications* **5**.

Kingma, D. and Ba, J. (2014), 'Adam: A method for stochastic optimization', *arXiv preprint arXiv:1412.6980* .

Kordás, K., Mustonen, T., Tóth, G., Jantunen, H., Lajunen, M., Soldano, C., Talapatra, S., Kar, S., Vajtai, R. and Ajayan, P. M. (2006), 'Inkjet printing of electrically conductive patterns of carbon nanotubes', *Small* **2**(8-9), 1021–1025.

Larson, C. (2018), 'Orbtouch'.
**URL:** *https://github.com/chrislarson1/orbtouch*

Larson, C., Peele, B., Li, S., Robinson, S., Totaro, M., Beccai, L., Mazzolai, B. and Shepherd, R. (2016), 'Highly stretchable electroluminescent skin for optical signaling and tactile sensing', *Science* **351**(6277), 1071–1074.

LeCun, Y., Bengio, Y. and Hinton, G. (2015), 'Deep learning', *Nature* **521**(7553), 436–444.

Lee, S.-S., Kim, S., Jin, B., Choi, E., Kim, B., Jia, X., Kim, D. and Lee, K.-p. (2010), How users manipulate deformable displays as input devices, *in* 'Proceedings of the SIGCHI Conference on Human Factors in Computing Systems', ACM, pp. 1647–1656.

Lipomi, D. J., Vosgueritchian, M., Tee, B. C., Hellstrom, S. L., Lee, J. A., Fox, C. H. and Bao, Z. (2011), 'Skin-like pressure and strain sensors based on transparent elastic films of carbon nanotubes', *Nature nanotechnology* **6**(12), 788–792.

Mikolov, T., Chen, K., Corrado, G. and Dean, J. (2013), 'Efficient estimation of word representations in vector space. arxiv'.

Mnih, V., Kavukcuoglu, K., Silver, D., Graves, A., Antonoglou, I., Wierstra, D. and Riedmiller, M. (2013), 'Playing atari with deep reinforcement learning', *arXiv preprint arXiv:1312.5602* .

Nakajima, K., Itoh, Y., Hayashi, Y., Ikeda, K., Fujita, K. and Onoye, T. (2013), Emoballoon: A balloon-shaped interface recognizing social touch interactions, *in* 'Virtual Reality (VR), 2013 IEEE', IEEE, pp. 1–4.

Nesterov, Y. (1983), 'A method of solving a convex programming problem with convergence rate o (1/k2)', *Soviet Mathematics Doklady* **27**(2), 372–376.

Ogata, M., Sugiura, Y., Makino, Y., Inami, M. and Imai, M. (2013), Senskin: adapting skin as a soft interface, *in* 'Proceedings of the 26th annual ACM symposium on User interface software and technology', ACM, pp. 539–544.

Ordóñez, F. J. and Roggen, D. (2016), 'Deep convolutional and lstm recurrent neural networks for multimodal wearable activity recognition', *Sensors* **16**(1), 115.

Pai, D. K., VanDerLoo, E. W., Sadhukhan, S. and Kry, P. G. (2005), The tango: A tangible tangoreceptive whole-hand human interface, *in* 'Eurohaptics Conference, 2005 and Symposium on Haptic Interfaces for Virtual Environment and Teleoperator Systems, 2005. World Haptics 2005. First Joint', IEEE, pp. 141–147.

Park, Y.-L., Majidi, C., Kramer, R., Bérard, P. and Wood, R. J. (2010), 'Hyperelastic pressure sensing with a liquid-embedded elastomer', *Journal of Micromechanics and Microengineering* **20**(12), 125029.

Rasmussen, M. K., Pedersen, E. W., Petersen, M. G. and Hornbæk, K. (2012), Shape-changing interfaces: a review of the design space and open research questions, *in* 'Proceedings of the SIGCHI Conference on Human Factors in Computing Systems', ACM, pp. 735–744.

Rogers, J. A., Someya, T. and Huang, Y. (2010), 'Materials and mechanics for stretchable electronics', *Science* **327**(5973), 1603–1607.

Roh, E., Hwang, B.-U., Kim, D., Kim, B.-Y. and Lee, N.-E. (2015), 'Stretchable, transparent, ultrasensitive, and patchable strain sensor for human–machine interfaces comprising a nanohybrid of carbon nanotubes and conductive elastomers', *ACS nano* **9**(6), 6252–6261.

Rus, D. and Tolley, M. (2015), 'Design, fabrication and control of soft robots', *Nature* **521**, 467–475.

Russomanno, A., OModhrain, S., Gillespie, R. B. and Rodger, M. W. (2015), 'Refreshing refreshable braille displays', *IEEE transactions on haptics* **8**(3), 287–297.

Shepherd, R., Peele, B., Murray, B. M., Barreiros, J., Shapira, O., Spjut, J. and Luebke, D. (2017), Stretchable transducers for kinesthetic interactions in virtual reality, *in* 'ACM SIGGRAPH 2017 Emerging Technologies', ACM, p. 21.

Stanley, A. A. and Okamura, A. M. (2017), 'Deformable model-based methods for shape control of a haptic jamming surface', *IEEE transactions on visualization and computer graphics* **23**(2), 1029–1041.

Steimle, J., Jordt, A. and Maes, P. (2013), Flexpad: highly flexible bending interactions for projected handheld displays, *in* 'Proceedings of the SIGCHI Conference on Human Factors in Computing Systems', ACM, pp. 237–246.

Stoppa, M. and Chiolerio, A. (2014), 'Wearable electronics and smart textiles: a critical review', *Sensors* **14**(7), 11957–11992.

Tang, S. K. and Tang, W. Y. (2010), Adaptive mouse: a deformable computer mouse achieving form-function synchronization, *in* 'CHI'10 Extended Abstracts on Human Factors in Computing Systems', ACM, pp. 2785–2792.

Viventi, J., Kim, D.-H., Vigeland, L., Frechette, E. S., Blanco, J. A., Kim, Y.-S., Avrin, A. E., Tiruvadi, V. R., Hwang, S.-W. and Vanleer, A. (2011), 'Flexible, foldable, actively multiplexed, high-density electrode array for mapping brain activity in vivo', *Nature neuroscience* **14**(12), 1599–1605.

Weigel, M., Mehta, V. and Steimle, J. (2014), More than touch: understanding how people use skin as an input surface for mobile computing, *in* 'Proceedings of the SIGCHI Conference on Human Factors in Computing Systems', ACM, pp. 179–188.

Yamada, T., Hayamizu, Y., Yamamoto, Y., Yomogida, Y., Izadi-Najafabadi, A., Futaba, D. N. and Hata, K. (2011), 'A stretchable carbon nanotube strain sensor for human-motion detection', *Nature nanotechnology* **6**(5), 296–301.